\documentclass[conference, 10pt]{IEEEtran}

\usepackage[
    letterpaper,
    paperwidth=215.9mm,
    paperheight=279.4mm, 
    left=13.2mm,
    right=13.2mm,
    top=22.9mm,
    bottom=25.4mm
]{geometry}

\usepackage{cite}
\usepackage{url}
\usepackage{graphicx}
\usepackage{color}
\usepackage{placeins}
\usepackage{float}
\usepackage{tabularx,colortbl}
\usepackage{ifthen}

\makeatletter

\newcounter{author}
\renewcommand{\author}[2][]{
   \stepcounter{author}
   \@namedef{author@\theauthor}{#2}
   \@namedef{authorlabel@\theauthor}{#1}
}

\newcounter{address}
\newcommand{\address}[2][]{
   \stepcounter{address}
   \@namedef{address@\theaddress}{#2}
   \@namedef{addresslabel@\theaddress}{#1}
}

\newcommand{\alsep}{and}

\def\newmaketitle{\par%
  \begingroup%
  \normalfont%
  \def\thefootnote{}
  \def\footnotemark{}
  \let\@makefnmark\relax
  \footnotesize
  \footnotesep 0.7\baselineskip
  \normalsize%
  \twocolumn[\thenewmaketitle\@IEEEaftertitletext]%
  \if@IEEEusingpubid
     \enlargethispage{-\@IEEEpubidpullup}%
  \fi
  \endgroup
  \setcounter{footnote}{0}\let\maketitle\relax\let\@maketitle\relax
  \gdef\@thanks{}%
  \let\thanks\relax}

\def\thenewmaketitle{
  \newpage
  \begin{center}%
    \vskip0.2em{\Huge\@IEEEcompsoconly{\sffamily}\@IEEEcompsocconfonly{\normalfont\normalsize\vskip 2\@IEEEnormalsizeunitybaselineskip
   \bfseries\large}\@title\par}\vskip1.0em\par%
    \vspace{1ex}
    \newcounter{c@author}
    \newcounter{c@tmp}
    \ifthenelse{\value{author}=2}{%
      \newcommand{\liand}{ and }}{%
      \newcommand{\liand}{, and }}
    \ifthenelse{\value{address}<2}{%
      \@nameuse{author@1}%
      \stepcounter{c@author}%
      \whiledo{\value{c@author}<\value{author}}{%
        \setcounter{c@tmp}{\value{author}}%
        \addtocounter{c@tmp}{-\value{c@author}}%
        \ifthenelse{\value{c@tmp}=1}{%
          \renewcommand{\alsep}{\liand}}{\renewcommand{\alsep}{, }}%
        \stepcounter{c@author}\alsep \@nameuse{author@\thec@author}}\\%
    }
    {
        \@nameuse{author@1}%
        \ifthenelse{\equal{\@nameuse{authorlabel@1}}{}}{}{${}^{(\ref{\@nameuse{authorlabel@1}})}$}%
        \stepcounter{c@author}%
        \whiledo{\value{c@author}<\value{author}}{%
          \setcounter{c@tmp}{\value{author}}%
          \addtocounter{c@tmp}{-\value{c@author}}%
          \ifthenelse{\value{c@tmp}=1}{%
            \renewcommand{\alsep}{\liand}}{\renewcommand{\alsep}{, }}%
          \stepcounter{c@author}\alsep \@nameuse{author@\thec@author}%
          \ifthenelse{\equal{\@nameuse{authorlabel@\thec@author}}{}}{}{${}^{(\ref{\@nameuse{authorlabel@\thec@author}})}$}%
        }
    }
    \vspace{0.2ex}

    \ifthenelse{\value{address}>0}{%
      \ifthenelse{\value{address}=1}{
        {\@nameuse{address@1}}
      }
      {
        \newcounter{c@address}

        \begin{center}
        \whiledo{\value{c@address}<\value{address}}
        {
          \refstepcounter{c@address}
            ${}^{(\thec@address)}$\,%
              \label{\@nameuse{addresslabel@\thec@address}}%
              \@nameuse{address@\thec@address}\\ %
        }
        \end{center}
      } 
    }
    {
      \relax
    }
  \end{center}
}

\makeatother

\title{Adjustment of Cluster-Then-Predict Framework for Multiport Scatterer Load Prediction}

\author[]{Hanjun Park$^{(\ref{org1})(\ref{org2})}$}
\author[org2]{Aleksandr D. Kuznetsov}
\author[org2]{Ville Viikari}

\address[org1]{Department of Electrical Engineering, Pohang University of Science and Technology, Pohang, South Korea}
\address[org2]{Department of Electronics and Nanoengineering, Aalto University, Espoo, Finland

parkterry@postech.ac.kr, \{aleksandr.kuznetsov,~ville.viikari\}@aalto.fi
}

\usepackage{balance}
\usepackage{booktabs}
\usepackage{amsmath} 
\usepackage{amssymb}
\usepackage{bm}
\usepackage[colorlinks=true, citecolor=black, linkcolor=black, urlcolor=blue]{hyperref}
\usepackage{siunitx}
\usepackage{multirow}
\usepackage{algorithmicx}
\newcounter{algorithm}
\usepackage{algpseudocode}
\usepackage{tikz} 
\ifCLASSOPTIONcompsoc
\usepackage[caption=false,font=normalsize,labelfont=sf,textfont=sf]{subfig}
\else
 \usepackage[caption=false,font=footnotesize]{subfig}
\fi

\begin{document}

\newmaketitle

\begin{abstract}
Predicting interdependent load values in multiport scatterers is challenging due to high dimensionality and complex dependence between impedance and scattering ability, yet this prediction remains crucial for the design of communication and measurement systems. In this paper, we propose a two-stage cluster-then-predict framework for multiple load values prediction task in multiport scatterers. 
The proposed cluster-then-predict approach effectively captures the underlying functional relation between S-parameters and corresponding load impedances, achieving up to a 46\% reduction in Root Mean Square Error (RMSE) compared to the baseline when applied to gradient boosting (GB). This improvement is consistent across various clustering and regression methods.
Furthermore, we introduce the Real-world Unified Index (RUI), a metric for quantitative analysis of trade-offs among multiple metrics with conflicting objectives and different scales, suitable for performance assessment in realistic scenarios. Based on RUI, the combination of K-means clustering and k-nearest neighbors (KNN) is identified as the optimal setup for the analyzed multiport scatterer.
\end{abstract}


\section{Introduction}
In recent years, the design of scatterers with engineered electromagnetic properties has attracted increasing attention due to their potential applications in various communication systems. Examples include reconfigurable intelligent surfaces (RISs) and backscattering devices \cite{lit_review_ris,backscat_inv_scat,our_S_param}.
One of the key approaches for designing such scatterers is the inverse scattering formulation \cite{backscat_inv_scat,optim_inv_scat,scatt_design_in_scat}.
Recently, numerous studies have proposed using artificial intelligence (AI) techniques to solve inverse scattering problems \cite{optim_inv_scat,scatt_design_in_scat,Phys_AI_inv_scat}, including real-time implementations \cite{realt_deepl_inv_scat} that are especially useful for the systems with dynamic update of parameters, such as RISs \cite{lit_review_ris}. 

As shown in \cite{our_S_param,Renzo_S_param_opt_fw_2024}, the terminations of loaded-antenna-based scatterers can strongly affect their scattering properties. This feature has been successfully exploited in RISs, where varying the load impedance values enables dynamic adaptation of scattering characteristics to the communication environment \cite{lit_review_ris}. Moreover, the design of impedance values in practical microwave circuits can be formulated within the inverse scattering framework \cite{imp_from_invscat}, which is also useful for advanced measurement techniques \cite{Virtual_VNA}.

\begin{figure}[!t]
\renewcommand{\figurename}{Algorithm}

\footnotesize 
\refstepcounter{algorithm}
Algorithm \thealgorithm. Training with Divide-and-Conquer Approach

\vspace{0.5ex}

\centering
\begin{tabular}{@{}l@{}}
\toprule
\begin{minipage}{0.95\columnwidth}
\begin{footnotesize}
\begin{algorithmic}[1]
    \Statex \textbf{Input:} Training dataset $D = \{X, Y\}$, No. of clusters $k$, No. of loads $L$
    \Statex \textbf{Output:} Clusters $C$, Regressors $M$
    \Function{Divide\_And\_Conquer}{$D, k, L$}
        \Statex{\color{gray}{\textbf{\textit{\qquad----- Divide -----}}}}
        \State $C \gets \text{Clustering}(D, k)$
        \State Initialize $k$ empty datasets (subset of $D$): $D_1, D_2, \cdots, D_k$
        \For{each $(x_i, y_i)$ $\in$ $D$}
            \State Find cluster $c_j$ that $x_i$ belongs to
            \State Add $(x_i, y_i)$ to dataset $D_j$
        \EndFor
        \Statex{\color{gray}{\textbf{\textit{\qquad----- Conquer -----}}}}
        \For{$j \gets 1$ to $k$}
            \State Initialize an empty regressor set $M_j$ for cluster $j$
            \For{$l \gets 1$ to $L$}
                \State $m_{jl} \gets \text{Train\_Regressor}(D_j, \text{load } l)$
                \State Add $m_{jl}$ to $M_j$
            \EndFor
            \State Add the set of regressors $M_j$ to $M$
        \EndFor
        \State \Return $C, M$
    \EndFunction
\end{algorithmic}
\end{footnotesize}
\end{minipage} \\
\bottomrule{
\label{alg1:divide_conquer}
}
\end{tabular}

\end{figure}

The integration of AI-based modeling and load adjustment for multiport scattering systems has been investigated in \cite{knn_gb_ours,NN_load_optimization}. These studies trained predictive models to estimate the required load terminations based on large datasets generated according to \cite{our_S_param}.
However, as reported in \cite{knn_gb_ours}, the prediction accuracy degrades notably in multi-load cases due to the high dimensionality of the data. A possible mitigation strategy is the \emph{cluster-then-predict} approach, where data are pre-clustered to reduce variability in sub-datasets before regression \cite{twostage_clusterpred_credit,twostage_clusterpred_build}.

In this work, we investigate the potential of a divide-and-conquer approach for predicting load impedance values in multiport scattering systems. 
Specifically, we apply the \emph{cluster-then-predict} approach combining k-nearest neighbors (KNN) and gradient boosting (GB) algorithms. 
The results demonstrate that prior clustering significantly improves the prediction accuracy of both algorithms that indicates the possible applicability of this approach for practical use in design of scattering systems.

\section{Theory}
\subsection{Task Description}

Consider a multiport scattering system terminated with impedance loads. Following \cite{our_S_param,Renzo_S_param_opt_fw_2024}, a strict connection exists between the terminated impedances and the scattered waves characteristics. Thus, predicting impedances from limited scattered-field data can be formulated as an inverse scattering problem where the target parameters are the load values. Similar task is also solved for ``virtual vector network analyzer (VNA)" \cite{Virtual_VNA}, which also could benefit from an alternative to the gradient-descent step.

In work \cite{knn_gb_ours}, this task was introduced as a first step toward optimization. The multi-load scenario, which allowed variation among load values, achieved poorer accuracy than the single-load case, especially for GB. 
This partially originates from preference of the algorithms for high-variance attributes, which suppresses weaker but relevant features. Dividing the dataset into smaller, more correlated subsets of S-parameters can mitigate this effect, improving prediction and reducing test time -- critical for real-time adjustable scatterers. With large synthetic datasets available, e.g., from \cite{our_S_param}, such improvement can be considerable due to high inter-sample variability, typical for S-parameters of scattering structures in different directions.
In this paper, for a direct comparison, we examine the use of prior clustering to create such sub-datasets, employing the similar dataset as in \cite{knn_gb_ours}.

\subsection{Cluster-then-Predict}

As demonstrated for different problem types \cite{twostage_clusterpred_credit,twostage_clusterpred_build,twostage_direction,twostage_clusterpred_lung}, a two-stage framework can improve performance when the analyzed data are large and complex. The underlying concept is to reduce computational complexity of the final algorithm by identifying the most relevant features in the dataset prior to training. One such method, the \emph{cluster-then-predict} approach, has been successfully applied in regression \cite{twostage_clusterpred_credit,twostage_clusterpred_build} and classification \cite{twostage_clusterpred_lung}.

To address the multi-load scenario, we adopt a non-recursive divide-and-conquer algorithm.
The training dataset is first divided into subsets through clustering, after which a regression model is trained for each cluster (Algorithm~\ref{alg1:divide_conquer}).
During inference, test samples are assigned to the nearest cluster via squared Euclidean distance to centroids (Algorithm \ref{alg2:cluster_predict}).
This two-stage \emph{cluster-then-predict} framework enables each regressor to operate on data subsets of smaller size and stronger internal correlation.

\begin{figure}[!t]
\renewcommand{\figurename}{Algorithm}

\footnotesize 
\refstepcounter{algorithm}
Algorithm \thealgorithm. Testing Phase: Cluster Assign and Prediction

\vspace{0.5ex}

\centering
\begin{tabular}{@{}l@{}}
\toprule
\begin{minipage}{0.95\columnwidth}
\begin{footnotesize}
\begin{algorithmic}[1]
    \Statex \textbf{Input:} $x_\text{new}$ $\in$ $X_\text{test}$, Clusters $C$, Regressors $M$, No. of loads $L$
    \Statex \textbf{Output:} Prediction vector  $\mathbf{y}_\text{pred} \in \mathbb{R}^L$
    \Function{Assign\_And\_Predict}{$x_\text{new}$, $C, M, L$}
        \State $c_\text{assigned} \gets \text{Assign\_To\_Cluster}(x_\text{new}, C)$
        \State $M_\text{regressor} \gets M[c_\text{assigned}]$
        
        \State Initialize $\mathbf{y}_\text{pred}$ as a zero vector of size $L$
        \For{$l \gets 1$ to $L$}
            \State $m_{\text{regressor}~l} \gets M_\text{regressor}[l]$
            \State $\mathbf{y}_\text{pred}[l] \gets m_{\text{regressor}~l}.\text{predict}(x_\text{new})$
        \EndFor
        
        \State \Return $\mathbf{y}_\text{pred}$
    \EndFunction
\end{algorithmic}
\end{footnotesize}
\end{minipage} \\
\bottomrule
\end{tabular}
{
\label{alg2:cluster_predict}}

\end{figure}

\subsection{Clustering via K-means and Optimal Transport}

We consider both standard K-means clustering \cite{K-means} and an Optimal Transport (OT) based K-means clustering variant \cite{otkmeans_ECCV}.
OT K-means clustering replaces the local assignment step with a global optimization yielding more accurate decompositions of data distributions, but at higher computational cost. 
For both methods, silhouette score \cite{silhouette_coeff} is used to evaluate cluster quality, with values ranging from $-1$ (overlapping) to $1$ (well-separated).

\subsection{Real-world Unified Index (RUI)}

For real-time applications such as RIS-supported channels and ``virtual VNA", both prediction accuracy and prediction time must be considered simultaneously. In the proposed two-stage framework, the quality of clusters is an additional factor to be considered.
To provide simultaneous evaluation under these multi-criteria conditions, we introduce the \emph{Real-world Unified Index} (RUI) -- a composite metric for scenarios:
\begin{enumerate}
    \item having multiple metrics
    \item with conflicting goals (e.g., minimizing Root Mean Square Error (RMSE) vs. maximizing silhouette score),
    \item and with different scales.
\end{enumerate}
RUI unifies these metrics into a single criterion indicating how optimal the selected option is for real-world implementation.

Consider a system characterized by $a+b$ metric values, where $a$ metrics $\mathbf{x}^{(i)}$ ($1\leq i \leq a$) should be maximized, and $b$ metrics $\mathbf{y}^{(j)}$ ($1\leq j \leq b$) should be minimized ($a, b \in \mathbb{N} \cup \{0\}$). Suppose that $m$ experiments are performed ($m\in\mathbb{N}$).
The first step in constructing RUI is normalization -- adjusting all metrics to the range $[0,1]$. 
Both groups of metrics $\mathbf{x}^{(i)},~\mathbf{y}^{(j)}~\in \mathbb{R}^{m \times 1}$ with conflicting goals are transformed to $\tilde{\mathbf{x}}^{(i)},~\tilde{\mathbf{y}}^{(j)}$ using:
\begin{equation}
\small
\tilde{\mathbf{x}}^{(i)}_k=\frac{\mathbf{x}^{(i)}_k-\min(\mathbf{x}^{(i)})}{\max(\mathbf{x}^{(i)})-\min(\mathbf{x}^{(i)})},
~\tilde{\mathbf{y}}^{(j)}_k=\frac{\max ( \mathbf{y}^{(j)} )-\mathbf{y}^{(j)}_k}{\max (\mathbf{y}^{(j)} )-\min(\mathbf{y}^{(j)})}~,
\end{equation}
where $1 \leq k \leq m$.
$\mathbf{RUI}\in \mathbb{R}^{m \times 1}$ is defined as:
\begin{equation}
\mathbf{RUI}=
\left[
\begin{array}{cccccccc}
| & & |&|&&|\\
\tilde{\mathbf{x}}^{(1)} & \cdots & \tilde{\mathbf{x}}^{(a)} & \tilde{\mathbf{y}}^{(1)} & \cdots & \tilde{\mathbf{y}}^{(b)}\\
|  & & |&|&&|\\
\end{array}
\right]
\begin{bmatrix}
w_1 \\
w_2 \\
\vdots \\
w_{a+b}
\end{bmatrix},
\label{eq2:rui-def}
\end{equation}
where $w_i~(1\leq i\leq a+b)$ is the weight for the $i$-th metric satisfying $\sum_{i=1}^{a+b} w_i = 1$. 
Weights are selected based on the importance of each metric for the considered real-world scenario.
\emph{Global} RUI enables comparison among methods by concatenating metric values from experiments across multiple methods for normalization, whereas restricting the evaluation to experiments within a single method defines \emph{local} RUI (or RUI$'$).
\footnote{The set of experiments of a \emph{single} method is not identical to $m=1$.}
Thus, identifying the method (or experiment) that maximizes \emph{global} (or \emph{local}) RUI for a given weight determines the \emph{optimal} choice.

\section{Experimental Setup}
Consider a multiport scattering system with three identical layers of coupled near half-wavelength dipoles of varying sizes, as described in \cite{our_S_param}.
To ensure compatibility, a similar dataset to that in \cite{knn_gb_ours} (specifically, Experiment 4 that targets the prediction of three load values) was utilized. 
The dataset consists of S-parameter data from four directions: $\mathrm{Re} \left(\frac{S_{\varrho,\tau}}{e^{i\alpha}} \right)$ for the intended scattering directions at $\phi=\SI{110}{\degree}, ~\SI{10}{\degree}$, and $|S_{\varrho,\tau}|$ for the parasitic scattering at $\phi= \SI{150}{\degree},~ \SI{60}{\degree}$. The required optimal phase shift $\alpha$ is set to 0 for $(\phi,~\theta)=(\SI{110}{\degree},~ \SI{90}{\degree})$ and $(\SI{10}{\degree},~\SI{90}{\degree})$.
A total of $N=10^6$ samples were split into an $80/20$ ratio for training and testing. The mean and standard deviation of each feature in the training set were used for normalizing both dataset parts.

The use of OT K-means is motivated by the toy example in \cite{optimaltransport_filip}, as our problem can be modeled as a mixture of four probability distributions where each S-parameter value approximately follows a normal distribution. 

To ensure a fair comparison, we used GPU (NVIDIA RTX A6000) based libraries for implementation: `cuML' for K-means clustering and KNN, `Python Optimal Transport (POT)' for OT K-means clustering, and `XGBoost' for GB. Silhouette score calculations were performed on CPU (AMD EPYC 7763 64-Core Processor).
Hyperparameters were set as: $\texttt{n\_init}=10$ for K-means; $\texttt{reg}=0.1$ for OT K-means; $\texttt{n\_neighbors}=5$ (with $3$ and $9$ for comparison) for KNN; and $\texttt{n\_estimators}=500$, $\texttt{learning\_rate}=0.3$, and $\texttt{max\_depth}=7$ for GB.

\section{Results and Discussion}

\begin{table}
\centering
\renewcommand{\arraystretch}{1.2}
\setlength{\tabcolsep}{2.5pt}
\caption{Performance Metrics corresponding to the configuration yielding the lowest RMSE for each method (3 Loads, 4 Angles)}
\begin{tabular}{l|c|c|cccc}
\toprule
Method& RMSE $\downarrow$ & Cluster. & $R^2$ $\uparrow$ & Sil. $\uparrow$& Train. [s]& Pred. [s]\\
\midrule
GB (Baseline) & 167.82 &- & 0.55 &-& 23.83 & 0.09\\
K-means + GB & \textbf{90.11} & 150 & 0.87 & 0.24 &552.40& 3.08\\
OT K-means + GB&92.39 & 160 & 0.86 & 0.15 & 1490.94 & 8.71\\
\cmidrule{1-7}
Maximum difference&\SI{-46}{\%}&-&$+$0.32&-&$+$\SI{1467}{s}&$+$\SI{9}{s}\\
\cmidrule{1-7}
KNN* (Baseline) & \textbf{82.36} &-& 0.89&-&1.28& 0.04\\
K-means + KNN* & 82.64 & 5 & 0.89 & 0.29 & 242.38 & 0.40\\
OT K-means + KNN*& 84.64 & 110 & 0.88 & 0.16 & 452.64 & 0.81\\
\cmidrule{1-7}
Maximum difference&$+$\SI{3}{\%}&-&$-$0.01&-&$+$\SI{451}{s}&$+$\SI{1}{s}\\
\bottomrule
\multicolumn{7}{l}{* $\texttt{n\_{neighbors}}=5$}\\
\multicolumn{7}{l}{Abbreviation ``Cluster.", ``Sil." stands for cluster size and silhouette score.}\\
\multicolumn{7}{l}{Baselines are the results without clustering. Maximum difference in respect}\\
\multicolumn{7}{l}{to the \textbf{baseline} is calculated.}\\

\end{tabular}
\label{tab1:best-acc-performance}
\end{table}
\subsection{The Effect of Clustering}

\begin{figure}[ht]
\centering
\subfloat[RMSE ($\downarrow$) values per cluster size (Cluster size 0 represents the \textbf{baseline})]{\includegraphics[width=1.0\columnwidth]{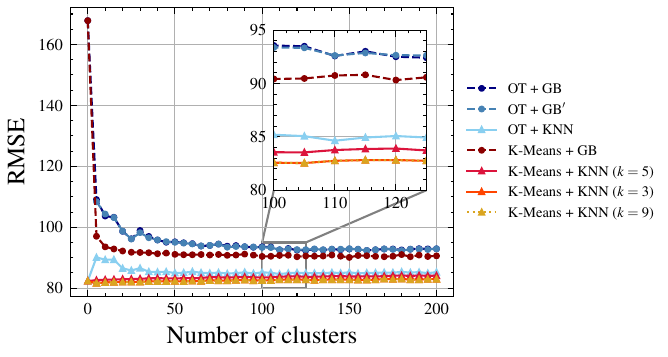}
\label{fig:per-cluster-(a)}}
\hfil
\subfloat[RUI ($\uparrow$) values per cluster size (Square markers represent local RUI maximum)]{\includegraphics[width=1.0\columnwidth]{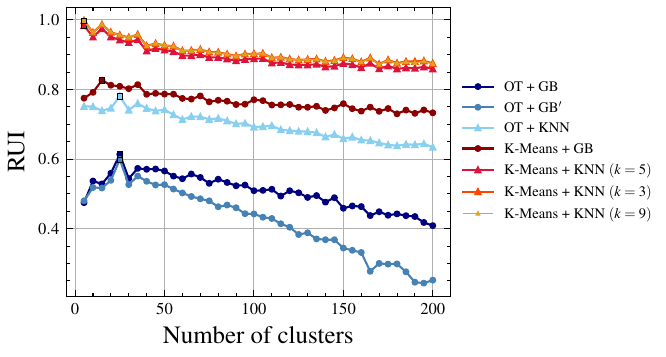}
\label{fig:per-cluster-(d)}}
\caption{Metric values per cluster size for 7 methods for 
K-means clustering (red-toned lines) and OT K-means clustering (blue-toned lines).
GB and KNN are distinguished using circle and triangle markers, respectively.}
\label{fig:per-cluster}
\end{figure}

To evaluate the effect of clustering, a grid-search on the number of clusters (from 5 to 200, step 5) was performed, and the accuracy (RMSE and $R^2$), silhouette score \cite{silhouette_coeff}, training time, and prediction time (time taken for cluster assignment and regression) were calculated for the following 7 methods: 
\begin{itemize}
    \item K-means $+$ \{GB, KNN $(k=3,~5,~9)$\}
    \item OT K-means $+$ \{GB, GB$'$, KNN $(k=5)$\}
\end{itemize}
GB$'$ uses single-load input/output, i.e., three regressors per cluster, whereas GB uses a multi-output model, i.e., one per cluster.
The upper bound for RMSE is benchmarked against the standard deviation of the test dataset (249.621~$\Omega$), while theoretical minimum is 0. 
A perfect $R^2$ score equals 1. 

As a result, we observe the following effects of clustering on RMSE, silhouette score, and prediction time.
First, as seen in Fig.~\ref{fig:per-cluster-(a)}, the two-stage  framework significantly improves GB, reducing RMSE until it plateaus.
This trend is also observed for $R^2$.
Conversely, KNN degrades slightly, suggesting clustering disrupts the local neighbor structure captured by the baseline~\cite{knn_gb_ours}. 
Second, K-means clustering yields higher clustering quality than OT K-means, though both exhibit declining silhouette scores as the number of clusters increases. 
This trend implies that excessive separation degrades quality by creating overlapping or unclear boundaries.
Also, we observed that GB prediction time scales linearly with the number of clusters. 
Furthermore, `OT $+$ GB$'$' is slower than `OT $+$ GB' because it involves three times as many models, a factor expected to lower its RUI. 
In contrast, KNN prediction time remains insensitive to cluster size, showing a slight decrease only when $\texttt{n\_neighbors}$ is reduced.

\subsection{Selecting the Optimal Cluster Size}
Table \ref{tab1:best-acc-performance} shows the metric values for the experiment that yields the \emph{lowest} RMSE for each method. 
The results indicate that the two-stage framework improves the RMSE of GB by \SI{46}{\%}, whereas KNN shows no significant improvement.
However, selecting the optimal cluster size should \emph{not} rely on accuracy (RMSE) alone.
Therefore, we utilize RUI, defined in Eq.~\eqref{eq2:rui-def}, to comprehensively consider multiple metrics.
In our setting, RUI consist of results concatenated across 7 methods (Fig.~\ref{fig:per-cluster}) with 40 different cluster sizes ($m=280$).  
Given $a=1$ (Silhouette score), $b=2$ (RMSE, Prediction time), and $m=280$, global RUI of $k$-th experiment is calculated as:
\begin{equation}
\mathbf{RUI}_k=0.3\times\mathrm{\tilde{\textbf{Sil}}}_k+0.4\times \tilde{\mathrm{\textbf{RMSE}}}_k+0.3\times\tilde{\mathrm{\textbf{Pred}}}_k,
\end{equation}
where $1\leq k \leq 280$. Since we prioritize the quality of clusters over time complexity, we set the weights to $\{0.3,~0.4,~0.3\}$.

For the method index $l \in \{ 0,1,\cdots, 6 \}$, local RUI is given by $\mathbf{RUI}'=\left \{ \mathbf{RUI}_i ~|~40l+1 \leq i \leq 40l+40 \right \}$.
The \emph{local} and \emph{global} optimal cluster sizes are chosen by computing $\underset{\text{cluster size}}{\mathrm{argmax}}~\mathbf{RUI}'$ and $\underset{\text{cluster size}} {\mathrm{argmax}}~\mathbf{RUI}$, respectively.
Fig.~\ref{fig:per-cluster-(d)} shows the local RUI maxima (square markers), revealing that KNN methods outperform GB and peak with fewer clusters.
\begin{table}
\centering
\renewcommand{\arraystretch}{1.2}
\scriptsize
\setlength{\tabcolsep}{2.5pt} 
\caption{Comparison of Local RUI Maxima across Experiments 
}
\begin{tabular}{llccc}
\toprule
Clustering & Regression & $\underset{\text{\{RMSE,Sil.,Pred.\}}}{\mathrm{argmax}} \mathbf{RUI}'$ & $\underset{\text{cluster size}}{\mathrm{argmax}}~\mathbf{RUI}'$ & $\mathrm{max}~ \mathbf{RUI}'$ \\
\midrule
K-means&GB&$\{$92.87, 0.28, 0.40$\}$&15&0.827\\ 
&KNN $(k=5)$&$\{$82.64, 0.29, 0.40$\}$&5&0.984\\ 
&KNN $(k=3)$&$\{$81.67, 0.29, 0.39$\}$&5&\textbf{0.998}\\ 
&KNN $(k=9)$&$\{$81.67, 0.29, 0.41$\}$&5&\textbf{0.998}\\ 
\cmidrule{1-5}
OT K-means&GB&$\{$96.21, 0.21, 1.64$\}$& 25&0.614\\
& GB$'$ &$\{$96.16, 0.21, 3.04$\}$& 25 & 0.599\\ 
&KNN $(k=5)$&$\{$85.82, 0.21, 0.25$\}$&25 &0.781\\ 
\bottomrule
\multicolumn{5}{l}{For KNN,  \texttt{n\_{neighbors}} is represented as $k$.}
\end{tabular}
\label{tab2:score-table}
\end{table}
Table~\ref{tab2:score-table} describes the configurations yielding the local RUI maximum for each method. 
We observe that `K-means $+$ KNN ($k=3, 9$)' achieves the global RUI maximum of 0.998, 
the highest value among all local RUI maxima. Although the two-stage framework is not effective for KNN as much as in GB, KNN is selected as the global optimal method, because it achieves relatively high accuracy even without clustering.

\subsection{Validating the Two-stage Cluster-then-predict Framework }
In addition, we validated the proposed two-stage \emph{cluster-then-predict} framework by extending it to various clustering (DBSCAN, Mean Shift) and regression models (Linear regression, Lasso, Linear SVR).
The framework yields consistent improvements in RMSE and $R^2$ across all models except for KNN-based frameworks (DBSCAN $+$ KNN, Mean Shift $+$ KNN), which shows negligible change (both giving RMSE 82.36 $\rightarrow$ 82.37, $R^2$ 0.89 $\rightarrow$ 0.89). 
While linear regression methods combined with OT K-means benefit significantly -- e.g., Linear regression (235.53 $\rightarrow$ 188.96, 0.11 $\rightarrow$ 0.42), Lasso (235.54 $\rightarrow$ 189.19, 0.11 $\rightarrow$ 0.42), and Linear SVR (238.01 $\rightarrow$ 205.52, 0.09 $\rightarrow$ 0.32) -- their absolute performance remains inferior to models like `DBSCAN $+$ GB' (167.82 $\rightarrow$ 113.19, 0.55 $\rightarrow$ 0.79) and `DBSCAN $+$ KNN'. 

\section{Conclusion}
This paper proposes \emph{cluster-then-predict} approach to improve load prediction accuracy in passive multiport scattering systems.
For a practical example of scattering system, the considered two-stage approach enhanced GB performance by \SI{46}{\%}. 
Using the proposed \textit{Real-world Unified Index}, the `K-means $+$ KNN' setup was identified as the most efficient, achieving a \SI{0.4}{s} prediction time.
Future work will focus on extending the \emph{cluster-then-predict} approach to classification problems in systems with a larger number of loads -- analogous to dynamic control in RISs -- and to the prediction of S-parameters from given load terminations.

\section*{Acknowledgment}

The research was partly funded by the WALLPAPER project of the Academy of Finland under decision 352913.
The authors acknowledge the use of MIDAS infrastructure of Aalto School of Electrical Engineering.

\bibliographystyle{IEEEtran}
\bibliography{References}

@ARTICLE{our_S_param,
  author={Kuznetsov, Aleksandr D. and Holopainen, Jari and Viikari, Ville},
  journal={IEEE Trans. Antennas Propag.}, 
  title={{Predicting the Bistatic Scattering of a Multiport Loaded Structure Under Arbitrary Excitation: The {S}-Parameters Approach}}, 
  year={2024},
  volume={72},
  number={8},
  pages={6691-6701},
  doi={10.1109/TAP.2024.3418517}
}

@article{lit_review_ris,
  author={Marco Di Renzo and Alessio Zappone and Merouane Debbah and Mohamed-Slim Alouini and Chau Yuen and Julien de Rosny and Sergei Tretyakov},
  journal={IEEE J. Sel. Areas Commun.}, 
  title={{Smart Radio Environments Empowered by Reconfigurable Intelligent Surfaces: How It Works, State of Research, and The Road Ahead}}, 
  year={2020},
  volume={38},
  number={11},
  pages={2450-2525},
  doi={10.1109/JSAC.2020.3007211}
}

@INPROCEEDINGS{NN_load_optimization,
  author={Kuznetsov, Aleksandr D. and Holopainen, Jari and Viikari, Ville},
  booktitle={2024 18th Eur. Conf. Antennas Propag. (EuCAP)}, 
  title={{Optimization of Loads for Antenna-Based Scattering Systems Using Feedforward Neural Networks}}, 
  year={2024},
  volume={},
  number={},
  pages={01-05},
  doi={10.23919/EuCAP60739.2024.10501204}
}

@ARTICLE{Renzo_S_param_opt_fw_2024,
  author={Abrardo, Andrea and Toccafondi, Alberto and Di Renzo, Marco},
  journal={IEEE Trans. Wireless Commun.}, 
  title={{Design of Reconfigurable Intelligent Surfaces by Using {S}-Parameter Multiport Network Theory—Optimization and Full-Wave Validation}}, 
  year={2024},
  volume={23},
  number={11},
  pages={17084-17102},
  doi={10.1109/TWC.2024.3450722}
}

@INPROCEEDINGS{knn_gb_ours,
  author={Kuznetsov, Aleksandr D. and Salmi, Albert and Holopainen, Jari and Viikari, Ville},
  booktitle={2025 19th Eur. Conf. Antennas Propag. (EuCAP)}, 
  title={{Capacity-Driven Smart Skin Loads Selection Utilizing {KNN} and Gradient Boosting}}, 
  year={2025},
  volume={},
  number={},
  pages={1-5},
  doi={10.23919/EuCAP63536.2025.10999591}}

@article{silhouette_coeff,
title = {{Silhouettes: A graphical aid to the interpretation and validation of cluster analysis}},
journal = {J. Comput. Appl. Math.},
volume = {20},
pages = {53-65},
year = {1987},
issn = {0377-0427},
doi = {https://doi.org/10.1016/0377-0427(87)90125-7},
author = {Peter J. Rousseeuw},
}

@ARTICLE{K-means,
  author={Lloyd, S.},
  journal={IEEE Trans. Inf. Theory}, 
  title={{Least squares quantization in {PCM}}}, 
  year={1982},
  volume={28},
  number={2},
  pages={129-137},
  doi={10.1109/TIT.1982.1056489}}

@ARTICLE{twostage_direction,
  author={Kim, Sanmin and Jeon, Hyeongseok and Choi, Jun Won and Kum, Dongsuk},
  journal={IEEE Robot. Autom. Lett.}, 
  title={{Diverse Multiple Trajectory Prediction Using a Two-Stage Prediction Network Trained With Lane Loss}}, 
  year={2023},
  volume={8},
  number={4},
  pages={2038-2045},
  doi={10.1109/LRA.2022.3231525}
}

@ARTICLE{twostage_clusterpred_lung,
  author={Leong, Christopher and Xiao, Yuanzhang and Yun, Zhengqing and Iskander, Magdy F.},
  journal={IEEE Access}, 
  title={{Non-Invasive Assessment of Lung Water Content Using Chest Patch {RF} Sensors: A Computer Study Using NIH Patients CT Scan Database and AI Classification Algorithms}}, 
  year={2023},
  volume={11},
  number={},
  pages={13058-13066},
  doi={10.1109/ACCESS.2023.3238969}}

@article{twostage_clusterpred_credit,
title = {{Bridging accuracy and interpretability: A rescaled cluster-then-predict approach for enhanced credit scoring}},
journal = {Int. Rev. Financ. Anal.},
volume = {91},
pages = {103005},
year = {2024},
issn = {1057-5219},
doi = {https://doi.org/10.1016/j.irfa.2023.103005},
author = {Huei-Wen Teng and Ming-Hsuan Kang and I-Han Lee and Le-Chi Bai}
}

@Article{twostage_clusterpred_build,
AUTHOR = {Ding, Zhikun and Wang, Zhan and Hu, Ting and Wang, Huilong},
TITLE = {{A Comprehensive Study on Integrating Clustering with Regression for Short-Term Forecasting of Building Energy Consumption: Case Study of a Green Building}},
JOURNAL = {Buildings},
VOLUME = {12},
YEAR = {2022},
NUMBER = {10},
ARTICLE-NUMBER = {1701},
ISSN = {2075-5309},
DOI = {10.3390/buildings12101701}
}

@ARTICLE{Phys_AI_inv_scat,
  author={Liu, Jian and Zhou, Huilin and Ouyang, Tao and Liu, Qiegen and Wang, Yuhao},
  journal={IEEE Trans. Antennas Propag.}, 
  title={{Physical Model-Inspired Deep Unrolling Network for Solving Nonlinear Inverse Scattering Problems}}, 
  year={2022},
  volume={70},
  number={2},
  pages={1236-1249},
  doi={10.1109/TAP.2021.3111281}
}

@INPROCEEDINGS{realt_deepl_inv_scat,
  author={Massa, Andrea and Chen, Xudong and Li, Maokun and Polo, Alessandro and Rosatti, Pietro and Salucci, Marco},
  booktitle={Proc. IEEE Int. Symp. Antennas Propag. USNC- URSI Radio Sci. Meeting (APS/URSI)}, 
  title={{Deep Learning: A Powerful Framework for the Real-Time Solution of Inverse Scattering Problems}}, 
  year={2021},
  volume={},
  number={},
  pages={2008-2009},
  doi={10.1109/APS/URSI47566.2021.9704174}
}

@INPROCEEDINGS{scatt_design_in_scat,
  author={Salucci, Marco and Hannan, Mohammad Abdul and Polo, Alessandro and Massa, Andrea},
  booktitle={Proc. IEEE Int. Symp. Antennas Propag. USNC- URSI Radio Sci. Meeting (APS/URSI)}, 
  title={{{AI}-Assisted Computationally-Efficient Global Optimization for Inverse Scattering}}, 
  year={2021},
  volume={},
  number={},
  pages={1687-1688},
  doi={10.1109/APS/URSI47566.2021.9704470}}

@ARTICLE{optim_inv_scat,
  author={Salucci, Marco and Poli, Lorenzo and Rocca, Paolo and Massa, Andrea},
  journal={IEEE Trans. Antennas Propag.}, 
  title={{Learned Global Optimization for Inverse Scattering Problems: Matching Global Search With Computational Efficiency}}, 
  year={2022},
  volume={70},
  number={8},
  pages={6240-6255},
  doi={10.1109/TAP.2021.3139627}
}

@ARTICLE{imp_from_invscat,
  author={de Padua Moreira, R. and de Menezes, L.R.A.X.},
  journal={IEEE Trans. Microw. Theory Tech.}, 
  title={{Direct synthesis of microwave filters using inverse scattering transmission-line matrix method}}, 
  year={2000},
  volume={48},
  number={12},
  pages={2271-2276},
  doi={10.1109/22.898974}}

@ARTICLE{backscat_inv_scat,
  author={Ma, Dingfei and Shen, Shanpu and Zhou, Hongxin and Zhang, Chi and Zhang, Qingfeng and Murch, Ross},
  journal={IEEE Trans. Antennas Propag.}, 
  title={{Integrated Sensing, Identification, and Backscatter Communication System Utilizing Inverse Scattering Approach}}, 
  year={2025},
  volume={73},
  number={8},
  pages={5877-5889},
  doi={10.1109/TAP.2025.3553972}}

@ARTICLE{optimaltransport_filip,
  author={Elvander, Filip and Haasler, Isabel},
  journal={IEEE Control Syst. Lett.}, 
  title={{Mixtures of Ensembles: System Separation and Identification via Optimal Transport}}, 
  year={2025},
  volume={9},
  number={},
  pages={1646-1651},
  doi={10.1109/LCSYS.2025.3581948}}

@InProceedings{otkmeans_ECCV,
author = {Mi, Liang and Zhang, Wen and Gu, Xianfeng and Wang, Yalin},
title = {{Variational Wasserstein Clustering}},
booktitle = {Proc. Eur. Conf. Comput. Vis. (ECCV)},
month = {September},
year = {2018},
pages={322-337}
}

@ARTICLE{Virtual_VNA,
  author={del Hougne, Philipp},
  journal={IEEE Trans. Instrum. Meas.}, 
  title={{Virtual {VNA}: Minimal-Ambiguity Scattering Matrix Estimation With a Fixed Set of “Virtual” Load-Tunable Ports}}, 
  year={2025},
  volume={74},
  number={},
  pages={1-19},
  doi={10.1109/TIM.2025.3551897}
}

\end{document}